\documentstyle[11pt]{article}

\def\kmu{$K^+\to \mu^+ \pi^0 \nu$}

\def\kp{$K^+\to \pi^+ \pi^0$}

\def\rkm{$K^+\to \mu^+ \nu \gamma$}
\def\kppn{$K^+\to \pi^+ \pi^0 \pi^0$}
\def\kpp{$K^+\to \pi^+ \pi^+ \pi^-$}
\begin{document}

\begin{center}

{\Large \bf Muon Polarization Working Group Report}

{ Presented at AGS2000 by Milind V. Diwan \\ Workshop on AGS experiments for the 21st Century \\
May12-17, 1996}

\vspace{4ex}

\small

Robert Adair$^1,$ 
Grigor Atoyan$^2,$
Basem Barakat$^3,$ 
Milind V. Diwan$^4,$

Jun Imazato$^5,$ 
Vladimir Issakov$^2,$
Yoshitaka Kuno$^5,$
Thaddeus Kycia$^4,$  
Richard Larsen$^1,$

Lawrence Leipuner$^4,$ 
Hong Ma$^4,$
Andrei Poblaguev$^2,$
Jack Ritchie$^6$

\medskip

{\it (1) Yale University, New Haven, Connecticut} \\
{\it (2) Institute for Nuclear Research, Moscow, Russia} \\
{\it (3) Louisiana Tech University, Ruston, Louisiana} \\
{\it (4) Brookhaven National Laboratory, Upton, New York} \\
{\it (5) KEK, National Laboratory for High Energy Physics, Japan} \\
{\it (6) University of Texas, Austin, Texas} 

\end{center}

\begin{abstract}
We have examined the physics and the experimental feasibility at the
AGS of various kaon decay processes in which the polarization of a
muon in the final state is measured.  Valuable information on CP
violation, the CKM matrix or new physics can be obtained with these
measurements and therefore they are well motivated.  In particular,
models of non-standard CP violation that produce the baryon asymmetry
of the universe could also produce effects observable in these
measurements. Limits from measurements such as the neutron and
electron electric dipole moment, and $\epsilon^\prime \over \epsilon$
in neutral kaon decays do not eliminate all of these models. We have
made a more detailed examination of the measurement of the out of the
plane muon polarization in \kmu ~decays.  With our current knowledge
of the AGS kaon beams and detector techniques it is possible to
measure this polarization with an error approaching $\sim 10^{-4}$.
Such an experiment would be well justified since the sensitivity is
well beyond the current direct experimental limit ($5.3\times
10^{-3}$) and the projected sensitivity ($< 10^{-3}$) of the currently
running experiment at KEK in Japan.
\end{abstract}

\vspace{4ex}

\section*{Introduction}

We have examined the possibility of measuring various muon 
decay asymmetries that are sensitive to 
P, T or CP symmetries; these are 
tabulated in Table \ref{list1}. 
Experimentally,  
CP violation has only been observed in  the neutral kaon system so far.
Although a  theoretical description of the 
CP-violation in the neutral kaon system exists through the complex phase 
in the Standard Model CKM matrix,  part or all of 
these phases could be consequences of 
deeper causes
that have so far eluded experiments.  
Over the last decade experiments 
at FNAL and CERN  directed towards the measurement of the direct 
$K^0_L \to \pi \pi$ transition or $\epsilon^\prime \over \epsilon$ have 
been inconclusive in  revealing the true nature of CP-violation.  Over 
the next decade ambitious efforts towards understanding CP-violation 
and the CKM matrix elements are planned with new $\epsilon^\prime
 \over \epsilon$
experiments and B-factories. The importance of these efforts is 
undeniable, yet
it must also be important to investigate the possibility that some or 
all of the 
CP-violation comes from effects outside the minimal Standard Model,  
particularly the CKM matrix.  

The CPT invariance of local quantum field theories requires
that CP violation is equivalent to T-violation. 
Therefore, it would be particularly interesting to look for 
direct violation of T-invariance outside the 
neutral kaon system.

It should also be noted that  CP-violation is required to generate  
the observed baryon asymmetry of the universe, and it 
is now accepted that  the CP-violation embodied in the CKM 
matrix does not have sufficient strength for this purpose [\ref{mclerran}]. 
Physics beyond the Standard Model that could generate the baryon 
asymmetry   can also generate 
CP or T violating muon polarizations in the kaon decay modes in Table \ref{list1}. 

\begin{table}
\begin{center}
\begin{tabular}{clccc}
\hline
\hline
 & Decay                &  Correlations & Symmetries \\
  &                     &              &  tested    \\
\hline
(1) & $K^+\to \pi^0 \mu^+ \nu$   &  $\vec s_\mu\cdot (\vec p_\mu\times \vec p_\pi)$ & T \\
\hline
(2) & $K^+\to  \mu^+ \nu \gamma$  &  $\vec s_\mu\cdot (\vec p_\mu\times \vec p_\gamma)$ & T \\
\hline
(3) &  $K_L\to  \mu^+ \mu^-$ &   $\vec s_\mu\cdot \vec p_\mu$ & P, CP \\
\hline
(4) &  $K^+\to \pi^+ \mu^+ \mu^-$ &  $\vec s_\mu\cdot \vec p_\mu$ & P \\
(5) &                             & $\vec s_\mu\cdot (\vec p_{\mu^+}\times \vec p_{\mu^-})$ & T \\
(6) &                             & ($\vec s_\mu \cdot \vec p_\mu) \vec s_\mu\cdot (\vec p_{\mu^+}\times \vec p_{\mu^-})$ & P, T \\
\hline
\hline 
\end{tabular}
\end{center}
\caption{\sl The decay modes and the polarization asymmetries 
or correlations of interest. }
\label{list1}
\end{table}

\begin{table}
\begin{center}
\begin{tabular}{llccccc}
\hline
\hline
Asym. & Mode & Branch. &  Standard   & Final      &  Non-SM    & Ref. \\
      &      & Fraction  &  Model    & State Int. &  value      &      \\
\hline 
(1) & $K^+\to \pi^0 \mu^+ \nu$ & 0.032 & 0.0 & $\sim 10^{-6}$ & $\le 10^{-3}$ & [\ref{garisto}] \\
\hline
(2) & $K^+\to \mu^+ \nu \gamma$ & $5\times 10^{-3}$ & 0.0 & $\sim 10^{-3}$ & $\le 10^{-3}$ & [\ref{kobayashi}] \\
\hline
(3) & $K_L \to \mu^+ \mu^-$ & $7\times 10^{-9}$ & $\sim 10^{-4}$ & 0.0 & $\le 10^{-2}$ & [\ref{gengng}, \ref{wolf1}] \\
\hline
(4) &   $K^+\to \pi^+ \mu^+ \mu-$ & $5\times 10^{-8}$  &  $\sim 10^{-2}$ & -- & -- & [\ref{wise1} --\ref{bgt}] \\
(5) &                &  & 0.0 & $\sim 10^{-3}$  & $\sim 10^{-3}$ & [\ref{anbg},\ref{gengsum}] \\
(6) &               &   & $\sim 6\times 10^{-2}$ & $\sim 0.0$ & $\sim 0.1$ & [\ref{anbg},\ref{gengsum}] \\
\hline
\end{tabular}
\end{center}
\caption{\sl The decay modes and asymmetries discussed by the working group.
The rest of the columns are:
the known branching ratio,
 the estimated Standard Model value,
the value due to final state interactions,
 the maximum possible
value allowed by non-standard  
physics, and the theoretical reference. }
\label{list2}
\end{table}

\section*{$K^+\to \pi^0 \mu^+ \nu$} 
The transverse or  out of plane muon polarization in this 
 decay has recently been analyzed by many authors [\ref{garisto},
\ref{belanger}, \ref{gengsum}]. The out of plane polarization is
expected to be zero to first order 
in the Standard Model because of the absence of 
the CKM phase in the decay amplitude. It has been shown that any 
arbitrary models involving effective V or A interactions cannot 
produce this type of polarization. Therefore, the existence of a non-zero
value of this polarization will be a definite signature of new 
physics beyond these models.  
In particular, some multi-Higgs  and 
leptoquark models could produce such a polarization. In multi-Higgs
models  a charged Higgs particle 
mediates an effective scalar interaction that interferes 
with the Standard Model decay amplitude; in such models 
the polarization could be as large as $10^{-3}$ without 
conflicting with other experimental constraints including 
the measurements of the neutron electric dipole moment and 
the branching fraction for $B\to X \tau \nu$, or $b\to s \gamma$ 
[\ref{alam}-\ref{aleph}].
 Irreducible  
backgrounds, i.e., final state interactions (FSI), to the 
out of plane polarization in this decay are expected to be 
small ($\sim 10^{-6}$) and therefore can be ignored [\ref{zitnitskii}].
 
The best previous experimental limits were obtained  almost 15 years 
ago with both 
neutral [\ref{schmidt}] and charged kaons [\ref{campbell}] at the BNL-AGS.
 The experiment with $K^+$ decays produced 
a measurement of the transverse polarization, $P^T_\mu = 0.0031 \pm 0.0053$.
The combination of both experiments could be interpreted as a limit on the 
 imaginary part of the ratio of the hadron form factors,
$Im\xi = Im(f_-/f_+) = -0.01 \pm 0.019$.
This limit is mostly independent of theoretical models 
and the experimental acceptance. 
By using the approximate formula, 
$P^T_\mu \approx 0.183 \times Im\xi$, one may reinterpret the above measure
of $Im\xi$ as a combined limit on the polarization, $P^T_\mu \approx 
-0.00185 \pm 0.0036$.  
This 1980 era measurement was based on  
$1.2\times 10^7 ~K^0_L$ 
and $2.1\times 10^7 ~ K^+$ decays to $\mu^+ \pi  \nu$ and was 
limited by statistics and backgrounds.  

Currently an experiment is in progress at the KEK-PS, E246 [\ref{kek246}], 
to measure $ P^T_\mu$ with a new technique of using 
a stopping $K^+$ beam and 
measuring the muon decay direction 
without spin precession. 
They expect to 
reach a sensitivity of $9\times 10^{-4}$ ($Im \xi < 4\times 10^{-3}$)
with $1.8\times 10^7$ events. 
The experiment will try to minimize systematics
by using the cylindrical symmetry of the apparatus and by using the
backward-forward $\pi^0$ 
symmetries of the decay at rest. The disadvantage of the stopping technique,
however,  is the low $K^+$ stopping rate. Nevertheless, the results of this
experiment will be very valuable to future experiments.

A new experiment has been designed 
at the BNL-AGS to perform this measurement
with an error on the polarization approaching $10^{-4}$
[\ref{newprop}].
 The design 
is based on  the 1980 experiment. 
The main improvement in the experiment will be the
 2 GeV/c separated charged kaon beam
decaying in flight. The separated beam will  reduce background 
counting rates in the detector per accepted event.
 The other improvements will be 
higher acceptance and analyzing power 
with a larger apparatus and a more finely 
divided polarimeter. 
Unlike the 1980 design the apparatus will also 
have better overall event reconstruction using tracking chambers and 
the larger calorimeter.
There is a possibility of improving the polarimeter design 
significantly using liquid scintillator mixtures that 
retain muon polarization [\ref{nakano}]. 
The experiment will collect approximately 550 events per AGS 
pulse per 3.6 seconds. Thus the statistical accuracy of the 
polarization measurement in a 2000 hr ($2\times 10^6$ pulses) run 
will be:
\begin{eqnarray}
\delta P_T \approx { 1.20^{1\over 2} 2^{1\over 2} \over 0.35
             (2\times 10^6 \cdot 550)^{1\over 2}  } 
            \approx 1.3\times 10^{-4} \nonumber
\end{eqnarray}
where $\sqrt 1.2 $,  $\sqrt 2$, $0.35$, are dilution factors in the 
analyzing power due to backgrounds, the precession magnetic field, 
and the muon 
decay, respectively.  The sensitivity to $Im{\xi}$ is given by 
$$\delta Im{\xi} \approx \delta P_T/0.2 \approx 7\times 10^{-4}$$
where 0.2 is a kinematic factor that includes the acceptance in 
the Dalitz plot and the orientation of the decay in the 
center of mass.  
With such high statistical power, systematic issues will become 
the main concern. The cylindrical symmetry of the apparatus 
and the precession technique (see [\ref{campbell}])
will cancel most systematic errors to first order. 
Nevertheless, the second order systematics will require some new 
techniques. The experiment will collect a large sample of data including 
$K^+ \to \mu^+ \nu$, $K^+ \to \pi^+ \pi^0$, and $K^+\to \pi^0 \mu \nu$
events in different parts of the decay phase space. The muon decay 
asymmetries from these various ensembles of events can be measured 
to understand the detector systematics to very high accuracy.

\section*{$K^+ \to \mu^+ \nu \gamma$}
The T violating out of plane polarization of the muon 
in this decay is related to the 
same  in $K^+\to \pi^0 \mu^+ \nu$ decay. The former 
can be caused by an effective pseudo-scalar interaction, while 
the latter  by an effective scalar interaction. Therefore searches
for T violation  
in both decay modes are complementary [\ref{kobayashi}].  
The T violating polarization could be  $\sim 10^{-3}$ 
without violating other experimental bounds. It is estimated [\ref{marciano}] 
that  the electro-magnetic 
FSI for this interaction can induce an out of plane muon polarization
of the same order of magnitude. An accurate theoretical calculation will 
be needed to subtract the FSI from any observation.
On the other hand, this FSI induced effect 
could be considered a useful calibration point for the apparatus that will 
also be used for the new $K^+ \to \pi^0 \mu^+ \nu$ experiment. 

The proposed new experiment  is optimized to study muon polarization in \kmu 
~decays.   Nevertheless, we have investigated the 
feasibility of measuring T-violation in \rkm.
The event selection and analysis of \rkm ~will be 
very similar to \kmu ~events except 
that events containing more than 1 photon 
 will be vetoed to reject background 
from \kmu, \kp, and \kppn ~events.
Further background rejection will be 
achieved by matching the measured muon 
 range in the 
polarimeter with the muon energy from a constrained fit to
the photon momentum, the muon direction, and the known 
kaon momentum. 
We expect to collect  $\sim 100$
events per AGS pulse. 
However,  the signal to background ratio  
with out current design will be 
about  0.3,  making it difficult to  reach sensitivities 
of 0.001 for the polarization.
Two improvements to the detector will reduce the backgrounds 
further:  
If the decay volume can be surrounded by photon veto counters with 
a veto threshold of 10 MeV
to detect the low energy photons from $\pi^0$ decays,
 the background level can be reduced to 
about 10\%. Secondly, if the calorimeter resolution can be improved 
(we have assumed $\sigma(E)/E \sim 8\%/\sqrt{E}$) then the 
muon range match can be made narrower, thus separating the
signal and background better.

\section*{$K_L\to \mu^+ \mu^-$}
The longitudinal muon polarization in this decay violates
CP invariance. This decay amplitude is known to be dominated by 
the two photon intermediate state. Interference of this amplitude 
with some other flavor changing neutral scalar interaction 
could produce a non-zero longitudinal polarization. Within 
the Standard Model such an interaction could take place through 
 second order loop diagrams involving the Higgs particle. However, 
direct constraints on the top quark and Higgs masses make the 
value of the polarization within the Standard Model  
quite small, $|{P_L(K_L\to \mu^+\mu^-)}|\sim 7\times 10^{-4}$ 
[\ref{gengng}].
Such a polarization could also arise in non-standard models that 
introduce new flavor changing neutral scalars. For example, 
Wolfenstein and Liu [\ref{wolf1}] have suggested 
that in two Higgs doublet models 
such a polarization could be as large as 0.10 without violating the 
bounds from the neutron electric dipole moment, $m_{K_L}-m_{K_S}$, $\epsilon$, and 
$\epsilon^\prime$.

The main experimental difficulty in this measurement is the 
small branching fraction of the decay, $7\times 10^{-9}$.
Therefore much effort must be put into separating these 
events from background before polarization analysis can be 
performed. Experiment E871 [\ref{871prop}] has collected the largest
number of these events so far; they expect to have $\sim 10000$ 
events at the end of the 1996 running period
with  little background. The experiment 
is optimized to look for $K_L \to \mu^\pm e^\mp$. We have made a
rough estimate that  
if the experiment  were optimized for 
$K_L\to \mu^+ \mu^-$ and the beam intensity were increased
E871 could collect about 20000 events in two 
years of running.  With appropriate upgrades to the marble 
muon range detector will allow approximately 50\% of the muon decays 
to be analyzed.
Y. Kuno has  suggested that a polarimeter   
made with  liquid scintillator could help this measurement by 
improving the analyzing power and 
lowering the cost of the polarimeter.
Thus aside from kinematic factors
the polarization could be measured with the following error:
\begin{eqnarray}
\delta P \approx {\sqrt{2}\over 0.3 \sqrt{10000}} \approx 0.05
\end{eqnarray}
where $\sqrt{2}$ and 0.3 are factors due to the precession magnetic
field in the polarimeter and the muon decay analyzing power, respectively.
We have not calculated 
the kinematic dilution factors that could
arise from the orientation of the decay in 
the center of mass of the kaon. 

\section*{$K^+\to \pi^+ \mu^+ \mu^-$}
This decay has recently been experimentally observed and measured 
to have the branching ratio of $5\times 10^{-8}$ [\ref{haggerty}]. 
The decay has a very rich structure which 
could lead to  important measurements: Table 
\ref{list1} shows three 
different  asymmetries that could be interesting to measure.
The decay has recently been analyzed quite extensively 
[\ref{wise1}--\ref{anbg}].
The several different processes that govern the decay are
as follows: one photon intermediate state, two photon intermediate
state, short distance graphs of ``Z-penguin'' and ``W-box'', and
potential contributions from extensions to the Higgs sector.
The interference of these graphs leads to various polarization effects. 
Although the theoretical analysis in the literature
does not seem to be complete -- in particular, strong interaction corrections 
and electro-magnetic final state effects  -- there is a consensus 
on the following: 

The CP conserving 
longitudinal polarization (asymmetry (4) from Table \ref{list1}) of 
the $\mu^+$ is sensitive to the Standard Model Wolfenstein parameter 
 $\rho$. The value of this polarization within the Standard Model 
is estimated to be $\sim 0.01$ and depends on the experimentally 
accepted phase 
space region. There is a small but non-negligible
 contribution to this 
polarization from the long distance 2 photon graph which cannot 
be calculated accurately at this time. 

The T violating out of plane polarization (asymmetry (5) in Table \ref{list1})
is expected to be very small within the Standard Model and the 
final state interaction correction to this polarization is expected to 
$\sim 10^{-3}$. T violating spin correlations that involve both $\mu^+$ and 
$\mu^-$ polarizations (asymmetry (6)) 
are expected to have much smaller final state 
interaction corrections and are theoretically  clean. Such asymmetries 
have substantial T violating contributions from the CKM matrix; they 
are expected to be $\sim 0.06$ in some parts of the 
decay phase space. A good measurement would be  sensitive to both the 
top quark mass and the CKM parameter $\eta$. It could also be 
sensitive to non-standard model physics in the same manner as 
$K_L \to \mu^+ \mu^-$.

Once again the main experimental difficulty will be in selecting 
the rare  $K^+\to \pi^+ \mu^+ \mu^-$ decays from background. The main 
background is \kpp ~decays in which the charged pions are misidentified 
as muons. This background must be suppressed in the  trigger and the 
analysis. Experiment E865 [\ref{865prop}] at the AGS is currently
  the best 
apparatus to perform this measurement. The experiment is, however, 
optimized for a search for $K^+ \to \pi^+ \mu^+ e^-$, and therefore 
will require some reconfiguration. In particular, the muon 
range finder will have to be changed to stop more muons and analyze 
the polarization. A rough estimate of the sensitivity can be made
based on the number of $K^+$ in the decay region of the experiment,
$\sim 10^7$ per AGS spill per 3.6 sec, the 
geometric acceptance for  $K^+\to \pi^+ \mu^+\mu^-$, 0.1, and the 
efficiency for muon decay detection of about 0.5. The longitudinal 
polarization could be measured 
in a 2000 hr run ($2\times 10^6$ spills) with an error of about:
\begin{eqnarray}
\delta P_L \approx {\sqrt{2}\over 0.3 (2\times 10^6\cdot 10^7\cdot 
5\times 10^{-8}\cdot 0.1 \cdot 0.5 )^{1\over 2} } \approx 0.02
\end{eqnarray}   
where $\sqrt{2}$ and 0.3 are factors due to the precession magnetic field,
and the muon decay analyzing power, respectively. 
Clearly, measuring asymmetries that require analyzing both $\mu^+$
and $\mu^-$ polarizations will be very difficult 
with current technology since $\mu^-$ 
decays have a much lower analyzing power due to muon capture 
into atomic orbits around
nuclei in the polarimeter. 

Table \ref{list1}  contains a summary of the various 
polarization asymmetries discussed by the working group. Table \ref{list2}
contains the approximate estimated values of the asymmetries 
within the Standard Model and outside the Standard Model. 
Some of the numbers from the various references 
have been adjusted to account for the new knowledge of the top
quark mass (174 $GeV/c^2$). In the case of $K_L\to \mu^+ \mu^-$ 
and $K^+ \to \pi^+ \mu^+ \mu^-$
the theoretical estimates for the maximum possible 
non-standard 
contributions to T violation do not agree; here the mean value of
various estimates is chosen.

\section*{Conclusion}

Muon polarization from kaon decays have a rich phenomenology.
In the case of $K_L \to \mu^+ \mu^-$ and 
$K^+ \to \pi^+ \mu^+ \mu^-$
new measurements could lead to important constraints on 
the Standard Model CKM parameters, in particular 
the Wolfenstein parameters $\rho$ and $\eta$. 
It is, however, difficult to reach the level of sensitivity needed 
to measure these parameters well with current technology. Nevertheless,
the experimental  difficulties should be compared to the difficulties
facing the rare kaon decay measurement of $K_L \to \pi^0 \nu \bar\nu$,
which  
 is sensitive to the same physics.

As shown in Tables 
\ref{list1} and \ref{list2}  
for many cases 
limits on the muon polarization will probe new physics 
beyond the Standard Model. In particular, 
the polarization will be sensitive to the physics of
a more complicated Higgs sector or leptoquarks that could 
give rise to CP or T violation outside the Standard Model.
The other source of CP violation needed for baryogenesis could be  
the motivation for such searches. 

We have examined the measurement of the out of plane 
muon polarization in \kmu ~decays in more detail. Such 
a measurement will not be sensitive to the Standard Model 
CP violation physics. 
Nevertheless, the measurement can be 
performed with sensitivity approaching $\delta P \sim 10^{-4}$, 
which is well beyond both the current direct limit 
of $\sim 5.3\times 10^{-3}$ and indirect 
 limit of $\sim 10^{-3}$ from 
other experimental constraints.
Although the electric dipole moments  of the
neutron and electron are considered more favorably
for T violation outside the Standard Model they do not 
cover the entire  spectrum of models. 
At the moment the measurement of T violating 
polarization in \kmu ~decays 
is well justified and should be considered complementary 
to other efforts in understanding CP violation.

We would like to thank Bill Marciano, Robert Garisto, Lawrence 
Littenberg, Steve Adler, Amarjit Soni for useful 
discussions.

\bigskip
\bigskip
\newpage

{\bf REFERENCES}

\begin{enumerate}

\vspace{-10pt}
\item \label{mclerran}
 L. Mclerran,  M. Shaposhnikov, N. Turok, and M. Voloshin, Phys. Lett. 
{\bf B 256}, 
451 (1991).  N. Turok and M. Voloshin, Phys. Lett. {\bf B 256}, 451 (1991). 
N. Turok and J. Zadrozny, Nucl. Phys. {\bf B 358}, 471 (1991). M. Dine, 
P. Huet,
R. Singleton, and L. Susskind, Phys. Lett. {\bf B 257}, 351 (1991). 
\vspace{-10pt} 
\item \label{garisto}
 R. Garisto and G. Kane, Phys. Rev. {\bf D 44}, 2038 (1991).
\vspace{-10pt}
\item \label{belanger}
 G.  Belanger and C. Q. Geng, Phys. Rev. {\bf D 44}, 2789 (1991).
\vspace{-10pt}
\item \label{alam} 
M.S. Alam, et al., Phys. Rev. Lett. {\bf 74}, 2885 (1995)
For the theoretical treatment in the context of the 3 Higgs
doublet model (3HDM) [\ref{weinberg}] see
Y. Grossman and Y. Nir, Phys. Lett. B {\bf 313}, 126 (1993).
\vspace{-10pt}
\item \label{aleph} 
D. Buskulic, et al., ALEPH collab., Phys. Lett.  {\bf B298}, 479 (1993).
For the theoretical treatment in the context of 3HDM see
Y. Grossman, 
Nuclear Physics {\bf B426}, 355 (1994).
\vspace{-20pt}
\item \label{zitnitskii}
A. R. Zhitnitskii, Sov. J. Nucl. Phys. {\bf 31}, 529 (1980).
\vspace{-10pt}
\item \label{schmidt}
 M. Schmidt, et al., Phys. Rev. Lett. {\bf 43}, 556 (1979).  W. Morse, et al., 
Phys. Rev.  {\bf D 21}, 1750 (1980).
\vspace{-10pt}
\item\label{campbell}
 M. Campbell, et al.,
 Phys. Rev. Lett. {\bf 47}, 1032 (1981). S. Blatt, et al., 
Phys. Rev.  {\bf D 27}, 1056 (1983).
\vspace{-10pt}
\item \label{kek246} 
J. Imazato, et al., KEK-PS research proposal Exp-246,
June 6, 1991.
\vspace{-10pt}
\item \label{kuno2}
Y. Kuno, Nucl. Phys. Proc. Suppl. {bf 37 A}, 87 (1993).
3rd KEK Topical Conference on CP Violation, Its Implications to Particle 
Physics and Cosmology, KEK, Tsukuba, Japan, Nov 16-18, 1993.
\vspace{-10pt}
\item \label{newprop}
A new proposal will be submitted to the AGS PAC in Oct. 96. 
\vspace{-10pt}
\item \label{nakano}
Takashi Nakano, Osaka University, Private Communication.
\vspace{-10pt}
\item \label{weinberg}
 S. Weinberg, Phys. Rev. Lett. {\bf 37}, 657 (1976).
\vspace{-10pt}
\item \label{kobayashi}
M. Kobayashi, T.-T. Lin and Y. Okada, Progress of Theoretical Physics, 
{\bf 95}, 261 (1996).
\vspace{-10pt}
\item \label{marciano}
W. Marciano, Private Communication.
See also C.Q. Geng, Nucl. Phys. B (Proc. Suppl.) {\bf 37A}
59 (1994). 
\vspace{-10pt}
\item \label{gengrkm2}
C.Q.Geng, Nucl. Phys. B {\bf 37A}, 59 (1994).
\vspace{-10pt}
\item \label{gengng}
F.J. Botella and C.S. Lim, Phys. Rev. Lett. 56 (1986) 1651.
Also see
C.Q. Geng, J.N. Ng, TRI-PP-90-64, Paper presented at the 
BNL CP summer study,
May 21-22, 1990.
\vspace{-10pt}
\item \label{wolf1}
J. Liu, L. Wolfenstein, Nuclear Physics, {\bf B289} 1 (1987).
\vspace{-10pt}
\item \label{871prop}
A. Heinson, et al., A New Search for Very Rare $K_L$ Decays,
AGS Proposal 871, Sep. 1990.
\vspace{-10pt}
\item \label{haggerty}
John Haggerty for the E787 collaboration,
 Proceedings of the XXVII International 
Conf. on High Energy Physics, GLASGOW, UK, JULY 20-27, 1994.
 edited by P.J. Bussey and 
I.G. Knowles. 
Published by Institute of Physics Publishing, Bristol and 
Philadelphia.
\vspace{-10pt}
\item \label{wise1} 
Ming Lu, Mark B. Wise, and Martin J. Savage, Phys. Rev. {\bf D46} 
5026 (1992).
\vspace{-10pt}
\item \label{wise2}
Martin J. Savage, Mark B. Wise, Phys. Lett. {\bf B250} 151 (1990).
\vspace{-10pt}
\item \label{gengsum}
C.Q. Geng, Talk presented at the KEK workshop on rare kaon 
decays, UdeM-LPN-TH-79, Dec. 10, 1991. 
\vspace{-10pt}
\item \label{buchalla}
G. Buchalla, A.J. Buras,  Phys. Lett. {\bf B336} 263 (1994).
\vspace{-10pt}
\item \label{gourdin}
Michel Gourdin, PAR-LPTHE-93-24, May 1993. 
\vspace{-10pt}
\item \label{kuno1}
 Yoshitaka Kuno (KEK, Tsukuba). 
KEK-PREPRINT-92-190, Jan 1993. 4pp. Published in
Proc. 10th Int. Symp. on High Energy Spin Physics, Nagoya, Japan, 
Nov 9-14, 1992. Page 769. 
\vspace{-10pt}
\item \label{bgt}
By G. Belanger, C.Q. Geng, P. Turcotte, 
Nucl. Phys. {\bf B390} 253 (1993).
\vspace{-10pt}
\item \label{anbg}
 Pankaj Agrawal, John N. Ng, G. Belanger, C.Q. Geng,
Phys. Rev. {\bf D45} 2383 (1992).
\vspace{-10pt}
\item \label{865prop}
E865 collaboration, BNL, PSI, Yale university, AGS  Proposal 865, May 7, 1990.
\vspace{-10pt}

\end{enumerate}

\end{document}